# Mesoscopic structural phase progression in photo-excited VO$_2$ revealed by time-resolved x-ray diffraction microscopy

*Yi Zhu[1], Zhonghou Cai[1], Pice Chen[2, †], Qingteng Zhang[2, §], Matthew J. Highland[3], Il Woong Jung[4], Donald A. Walko[1], Eric M. Dufresne[1], Jaewoo Jeong[5], Mahesh G. Samant[5], Stuart S. P. Parkin[5,6], John W. Freeland[1], Paul G. Evans[2], Haidan Wen[1*]*

[1] Advanced Photon Source, Argonne National Laboratory, Argonne, Illinois 60439, USA
[2] Department of Materials Science and Engineering, University of Wisconsin–Madison, Madison, Wisconsin 53706, USA
[3] Materials Science Division, Argonne National Laboratory, Argonne, Illinois 60439, USA
[4] Center for Nanoscale Materials, Argonne National Laboratory, Argonne, Illinois 60439, USA
[5] IBM Almaden Research Center, San Jose, California 95120, USA
[6] Max Planck Institute for Microstructure Physics, Halle 06120, Germany

*Corresponding author: wen@aps.anl.gov

[†] Present address: Department of Materials Science and Engineering, Northwestern University, Evanston, Illinois 60208, USA

[§] Present address: Advanced Photon Source, Argonne National Laboratory, Lemont, Illinois 60439, USA

**Dynamical phase separation during a solid-solid phase transition poses a challenge for understanding the fundamental processes in correlated materials. Critical information underlying a phase transition, such as localized phase competition, is difficult to reveal by measurements that are spatially averaged over many phase separated regions. The ability to simultaneously track the spatial and temporal evolution of such systems is essential to understanding mesoscopic processes during a phase transition. Using state-of-the-art time-resolved hard x-ray diffraction microscopy, we directly visualize the structural phase progression in a $VO_2$ film upon photoexcitation. Following a homogenous in-plane optical excitation, the phase transformation is initiated at discrete sites and completed by the growth of one lattice structure into the other, instead of a simultaneous isotropic lattice symmetry change. The time-dependent x-ray diffraction spatial maps show that the in-plane phase progression in laser-superheated $VO_2$ is via a displacive lattice transformation as a result of relaxation from an excited monoclinic phase into a rutile phase. The speed of the phase front progression is quantitatively measured, and is faster than the process driven by in-plane thermal diffusion but slower than the sound speed in $VO_2$. The direct visualization of localized structural changes in the time domain opens a new avenue to study mesoscopic processes in driven systems.**

Phenomena such as insulator-to-metal phase transitions (IMT), colossal magnetoresistance, and high-$T_c$ superconductivity are often characterized by spontaneous appearance of spatially separated regions with distinct structural, magnetic, and electronic properties[1]. Such phase heterogeneities span length scales from nanometers to micrometers and have roles in determining dynamics of phase transformations, electron transport, and magnetic and structural responses to applied fields[2–7]. Among these phenomena, the IMT has broad fundamental interest and technological relevance. For example, as an archetypical correlated electron material, vanadium dioxide ($VO_2$) exhibits an orders-of-magnitude change of the electrical conductivity across the close-to-room-temperature transition, accompanied by a change in lattice symmetry[8,9]. This transition can be triggered by optical excitation on ultrafast time scales, holding promise for low-power ultrafast electronics[10]. The ultrafast nature of the transition is also of fundamental interest because the electronic and structural degrees of freedom can be disentangled in the time domain to help resolve a decades-long debate concerning the mechanisms of the phase transition in $VO_2$[11]. Extensive studies of the initial sub-picosecond (ps) effects by ultrafast optical, x-ray, and electron probes have yielded significant mechanistic insights of IMT[12–24]. New discoveries such as the observation of a metal-like monoclinc phase[23] and inhomogenous onset transition time scales[18] stimulates futher microscopic investigation on the phase transition in $VO_2$.

On longer time scales, the picture of a uniform transformation across the optically excited region is challenged by the complex interactions among spatially separated phases arising from structure disorder and elastic coupling[25–27]. Characterizing the phase transformation processes from the initiation to the completion of photoinduced phase transition is essential not only to understanding the fundamental physics of mesoscale phenomena, but also to harnessing IMTs for device applications. Early studies in the time[28–31] and space[18,32–37] domains have shown the dynamical and heterogeneous nature of the phase transition respectively; however, the microscopic processes including spatial progression of the phase transition, the energetics and dynamics of phase boundaries, and the characteristic length scale and speed of the transformation, are not yet known. Quantitative characterization of these mesoscopic processes requires the visualization of the dynamical processes in appropriate time and space domains simultaneously.

Using a newly developed laser pumped x-ray diffraction imaging technique with 350 nm spatial resolution and 100 ps temporal resolution[38], we quantitatively studied the structural phase propagation during the photo-induced phase transition in a $VO_2$ thin film (Fig. 1a). Time-resolved mapping of the real-space structural transformation (Fig. 1b) is captured by tracking signatures of the structural phases in the reciprocal space (Fig. 1c). Following the initial ultrafast photo-induced structural phase transition[19–23], the optical excitation creates an unstable high-energy monoclinic (M) state, where the energy supplying the latent heat has already been stored in the system before the transition occurs. This high energy state transitions to the rutile (R) state via a displacive lattice symmetry change, i.e., a cooperative motion of atoms, in a superheated state of $VO_2$. The resulting high-temperature R phase propagates from a series of nucleation sites into regions of the M phase (Fig. 1b). This process is not driven by thermal diffusion since the quantitative measurement of the phase progression speed is faster than that is predicted by thermal diffusion. This experiment represents the first hard x-ray measurement with sub-ns and sub-µm resolution that directly captures a mesoscopic structural phase transformation in correlated materials.

**Results**

Steady-state diffraction measurements show distinct diffraction of the $40\bar{2}$ Bragg peak in the low-temperature M phase and the 002 Bragg peak in the high-temperature R phase in Fig. 2a. The angular widths of the two Bragg peaks are sufficiently broad so that both x-ray reflections can be simultaneously measured at the appropriately chosen x-ray incident angle, with separated diffraction peaks recorded by an x-ray photon-counting area detector (Fig. 2b).

For the time-resolved experiments reported below, the x-ray incident angle was set to the position corresponding to the peak intensity of the M phase and the low-angle tail of the R phase, as indicated by the vertical dashed line in Fig. 2a.

The time-resolved measurement over a large area using a 50 µm full-width half maximum (FWHM) x-ray probe beam shows two distinct time scales during the phase transformation (Fig. 2c). The sub-ns change in diffraction intensity is dominantly driven by the out-of-plane phase progression over tens of nm length scales as heat flows inwards from the top excited layer of the $VO_2$ film. The change on ns time scales is associated with the in-plane displacive phase progression over µm length scales.

We first examined the propagation of the phase transformation along the out-of-plane direction. Since the $VO_2$ film is 200 nm thick, twice the penetration depth of an 800 nm excitation pulse[39], the optical excitation along the out-of-plane direction is depth-dependent while the x-ray diffraction probes through the whole depth of the film[40]. At $t_1$=100 ps after optical excitation with a fluence of 14 mJ/cm$^2$, the diffraction intensity from M phase decreased as a fraction of the excited region was transformed into R phase, leading to the increase of the diffraction intensity from R phase, as shown in Fig. 2c. The quantitative analysis shows that the subsequent phase front propagation of ~50 nm in 1 ns along the out-of-plane direction was driven by thermal transport as heat diffused into the deeper layers (see Supplemental Materials), illustrated by Fig. 2d. This independent characterization of thermally driven dynamics in the same sample provides a direct reference to compare with the dynamics of the in-plane displacive structure phase progression.

Although the in-plane phase progression occurs simultaneously at sub-ns time scales, its dynamics cannot be resolved in the measurement with the 50 µm x-ray probe beam. The quantitative analysis of the in-plane phase progression requires x-ray probes with orders of magnitude higher spatial resolution. The existence of ns-timescale dynamics following laterally homogeneous optical excitation suggests that the completion of the phase transformation occurs not within a few ps isotropically but across the sample on mesoscopic time (ns) and length (μm) scales inhomogenously.

To study the phase progression along in-plane direction, time-dependent spatial maps of the M and R phase diffraction intensity were collected by raster scanning the sample against a focused x-ray probe beam at various delays (Fig. 3). The focused x-ray beam may alter the electronic properties of $VO_2$[41], but no structural modification of the sample was observed during our measurements. The average diffraction intensities across the entire $10 \times 10$ μm$^2$ maps are shown as the open squares and circles in Fig. 2c, consistent with the measurements where the film was probed using the 50 μm beam (filled symbols). Since the lattice structure was monoclinic before the optical excitation, at $t = -0.5$ ns, the x-ray diffraction exhibited a high M-phase diffraction intensity map while the R-phase map showed no diffraction intensity above the noise level.

After optical excitation, the R phase diffraction intensity increases initially at discrete locations, e.g., site A in Fig. 3. These locations are associated with the initial photo-induced R phase nuclei, which coalesce to form R-phase sites observable with a 350 nm x-ray probe beam. In the context of the overall transformation, the initial photoinduced phase transformation creates nucleation sites for the subsequent growth of R phase. The overall transition requires the slower transformation of the remaining $VO_2$ thin film. We note that the time-resolved images are obtained by averaging over many pump-probe experimental cycles, which capture the emergence of R phase sites that follow a deterministic transition pathway in each pump-probe cycle.

The propagation speed of the R-M phase front can be quantitatively characterized. A line-cut of the R phase diffraction intensity through site A, shown as the red dashed line in Fig. 3, is plotted as a function of time in Fig. 4a. The R phase expands laterally following its initial appearance at site A with a velocity of $1100 \pm 300$ m/s, which is extracted from the slope of the dashed black line in Fig. 4a. The experimental error of the propagation speed reflects the uncertainty in determining the phase boundary, which is defined as the 50%-intensity position in Fig. 4a. At this speed, it only takes 45 ps for the phase boundary to progress 50 nm. This time scale is much shorter than 1 ns needed for the thermal diffusion-driven phase progression, as independently measured along the out-of-plane direction in the same sample. Similar analysis at other sites yields velocities ranging from several hundred to 1100 m/s, lower than the sound velocity of 4 km/s in $VO_2$[42], as indicated by the solid black line in Fig. 4a. The slower phase progression than sound wave propagation may be a result of grain boundaries and local strain states that impede phase progression[35–37,43].

The spatial progression of phases is further supported by the location-dependent dynamics of the photo-induced phase transition. We plotted the diffraction intensities as a function of delay for three regions of interests: A, B, and C in Fig. 4b. The evolution of the R phase at these sites are different after the optical excitation. For example, the phase transformation at site B required a longer time (2.5 ns) to reach the same R phase composition at site A, which transforms on a far shorter time scale (0.7 ns). The R phase composition at site C remained lower than site A and B in ~3 ns, the longest delay time probed in these maps. These observations show that the site A served as a nucleation site while the transition at site B and C occurred at later time since the transition of these sites needs to wait for the arrival of the phase transition front.

**Discussion**

The visualization of an inhomogeneous solid-solid phase transformation provides insight into the multi-stage processes for completing the phase transformation. The first stage is the initial symmetry change at nucleation sites on the unit-cell level on fs time scales following optical excitation as reported previously[19–23]. These nucleation sites are possibly related to local strain states[35–37] in favor of the stabilization of the high-temperature phase, and variations in grain size because smaller single crystalline particles require less activation energy[31]. The second stage, following the initial fs phase transition, is a process in which the lattice is superheated above the transition temperature within a few ps, as a result of electron-phonon coupling during which the absorbed radiation energy in electronic degree of freedom is transferred to the lattice. At this stage, except at the nucleation sites, the film enters a transient state with sufficiently high energy for the structural phase transition as a result of optical absorption, but a lattice symmetry change has not yet occurred. The third stage is a domino-like lattice transformation propagating from the nucleation sites into the excited but not yet transformed $VO_2$ film. This process is a displacive lattice transformation[44], in which the lattice symmetry change is a result of straightening V-V atom pairs without diffusive heat transport, similar to the crystallization or melting which starts from shear strain in supercooled or superheated solids[45–47]. The displacive lattice transformation is supported by the fast phase progression speed that exceeds the thermal diffusion limit.

We note that the homogenous ultrafast excitation results in little variation of temperature in the probed region. Therefore, the in-plane phase progression in ultrafast superheated $VO_2$ does not rely on thermal transport in which temperature gradients and thermal diffusion effects drive the phase progression. Instead, the transformation is a result of stabilizing the lattice by lowering the superheated lattice energy. The excessive energy absorbed during optical excitation is consumed as the latent heat needed for the first-order phase transition. Therefore, the formation of R phase islands does not create temperature gradients since the latent heat has been supplied by optical absorption. The transition from M to R phase is followed by a slow recovery from R to M phase on hundreds of ns time scales as the film cools down via heat transport into the substrate.

The observed R-phase pinning sites are less numerous than the expected defect sites. We observed several pinning sites in the $10 \times 10$ μm$^2$ maps, whereas the density of grain boundaries in the thin films is orders of magnitude larger[37]. The emergence of R phase at site A is a result of growing R-phase domains via coalesce of smaller nucleation sites. Between the observed pinning sites, there could be many smaller domains that cannot be resolved in our measurements. These domains do not coalesce to form larger detectable domains if the absorbed energy is not sufficient to overcome local energy barrier across domain boundaries[35].

In conclusion, we have directly characterized the structural evolution of a photo-induced phase transformation in $VO_2$ using time-resolved x-ray diffraction microscopy. The sub-µm length and sub-ns time resolution allows us to reveal the structural phase progression during the photoinduced phase transition. The inhomogeneous structural phase transformation that develops from nucleation sites can be understood by a displacive lattice motion in the superheated $VO_2$, which evolves from the unstable high-energy M phase into a metastable low-energy R phase. Further studies using the same technique on different $VO_2$ thin films will reveal, for example, the dependence of phase progression speed on the grain size and crystalline orientations. This new aspect of the structural phase transition is important not only for understanding mesoscale phase transformations but also for designing ultrafast optoelectronic devices using $VO_2$. For example, faster structural phase transformation as a whole may occur in $VO_2$ with more nucleation sites, while faster phase front progression that approaches the sound speed may occur in ultrathin $VO_2$ films that contain fewer grain boundaries. The visualization of phase transformations with time-resolved x-ray imaging technique opens new opportunities to study nanoscale phase transition and separation in correlated materials in the time domain.

**Methods**

***Time-resolved synchrotron hard x-ray diffraction microscopy.*** The experimental setup at the 7-ID-C beamline of the Advanced Photon Source is shown schematically in Fig. 1a and the technical details are provided in the Supplemental Materials. Laser pulses with 1.55 eV photon energy and 60 fs pulse duration were focused to a 200 µm (FWHM) spot to excite the $VO_2$ film. The 1 kHz laser repetition rate allowed full recovery of the excited $VO_2$ to the ground state between excitation pulses. At an incident fluence of 14 mJ/cm$^2$, the excited VO$_2$ film was superheated to a temperature of $T$~380 K at 100 ps after excitation, as estimated by the absorbed energy and thermal properties of VO$_2$[29]. The 10 keV hard x-ray pulses were focused by a Fresnel zone plate to a spot size of 350 nm (FWHM), significantly smaller than the pump beam. The time delay between the optical and x-ray pulses was electronically adjusted, with the temporal resolution of 100 ps set by the duration of the x-ray pulses[48].

*Sample preparation.* A 200 nm-thick VO$_2$ film was grown by pulsed-laser-deposition on an Al$_2$O$_3$ ($10\bar{1}0$) substrate at 700 °C, with a deposition laser fluence of ~1.3 J/cm$^2$ and repetition rate of 2 Hz.

**Acknowledgements**

We thank Michael J. Wojcik for manufacturing the Fresnel zone plate. We appreciate the insightful discussion with Aaron Sternbach and Dimitri N. Basov. Work at Argonne was supported by the U.S Department of Energy, Office of Science, Office of Basic Energy Sciences, under Contract No. DE-AC02-06CH11357. H. W. and J. F. acknowledge the support for the data analysis from the U.S Department of Energy, Office of Science, Office of Basic Energy Sciences, under Contract No. DE-SC0012375. Work at the University of Wisconsin was supported by the U.S. Department of Energy, Office of Basic Energy Sciences, Division of Materials Sciences and Engineering, through Grant No. DEFG02-10ER46147.

**Author contributions**

H. W. conceived the experiments. Y. Z., Z. C., P. C., Q. Z., M. H., I. J., D. W., E. D., J. F., P. E. and H. W. designed and performed the measurements. J. J., M. S., and S. P. grew the sample. Y. Z. and H. W. wrote the paper with contributions from all authors.

**Additional information**

Competing financial interests: The author(s) declare no competing financial interests.

**Figures and Captions**

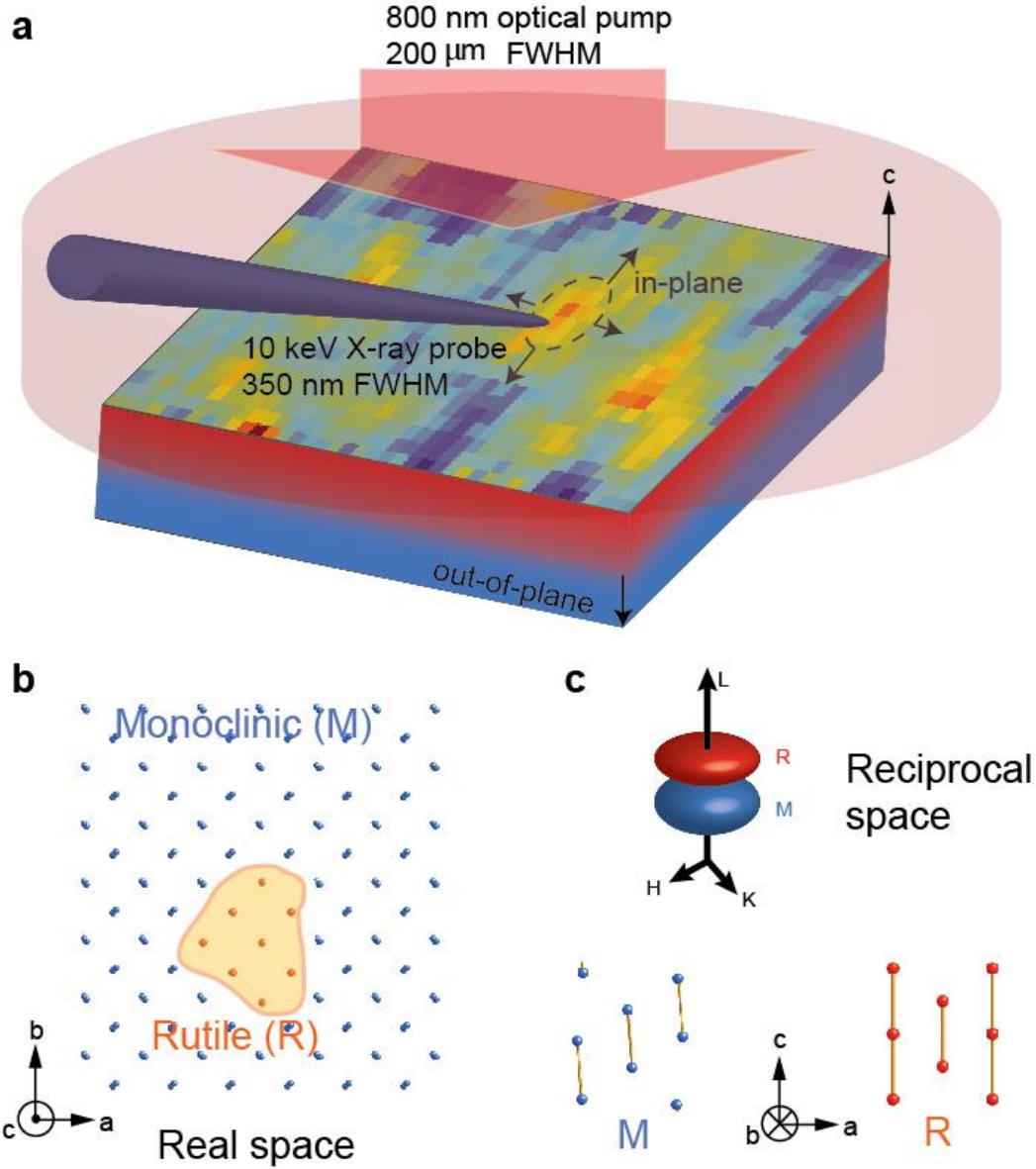

**Figure 1** (a) Laser-pumped hard x-ray diffraction microscopy. The structural phase progression of a $VO_2$ film along in-plane and out-of-plane directions, indicated by the black arrows, is probed by synchrotron-based focused x-ray pulses upon homogenous optical excitation along the in-plane direction. The blue and red regions represent monoclinic (M) and rutile (R) phases respectively. (b) A schematic of the real space arrangement of atoms (only V atoms are shown). The contour highlights the M-R phase boundaries during phase transformation. The *a,b,c-* axes are labeled in the R coordinates. (c) The reciprocal space map of the diffraction patterns of R and M phases and the corresponding lattice structures.

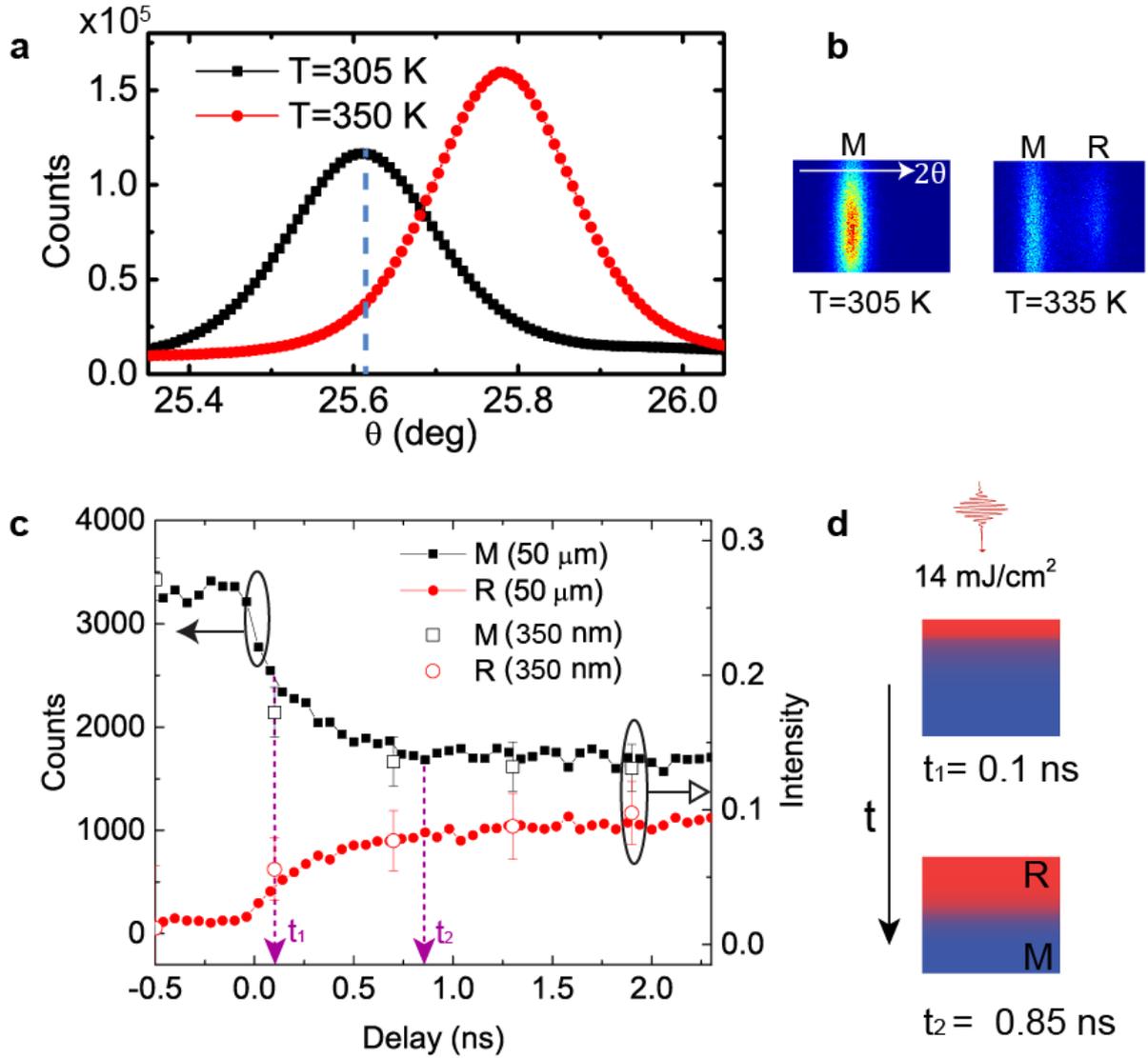

**Figure 2** (a) The $40\bar{2}$ M phase (T=305 K) and 002 R phase (T=350 K) Bragg reflections measured by 10 keV x-ray radiation. The blue dashed line indicates the incident x-ray Bragg angle $\theta=25.62°$ at which the time-resolved measurements are performed. (b) The M and R diffraction images are measured below and at the transition temperatures by an x-ray area detector. The white arrow points to higher $2\theta$ direction. (c) Diffraction intensities of the M and R phases as a function of delay, measured with a 50 μm (FWHM) x-ray beam (filled symbols) and with a 350 nm x-ray beam (open symbols, see text) at a fluence of 14 mJ/cm$^2$. The open symbols show the averaged diffraction intensity of the 2D maps in Fig. 3 normalized by a detector reading which is proportional to the incident x-ray flux. The purple arrows show the start ($t_1$=100 ps, limited by x-ray pulse duration) and the end ($t_2$, see Supplementary Materials) time of the out-of-plane phase progression. (d) Schematics of the out-of-plane progression at two delays. The red and blue regions represent R and M phases respectively.

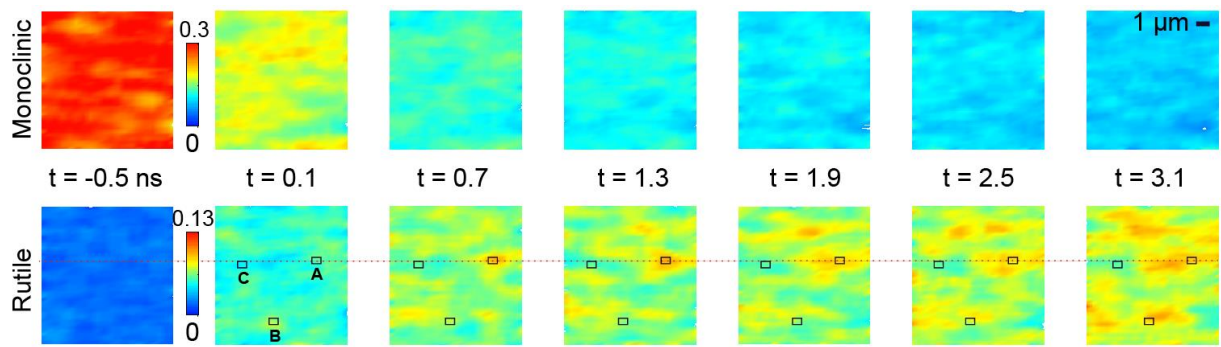

**Figure 3** The intensity maps of the M and R phases measured at a sequence of time delays excited by an optical pulse with a fluence of 14 mJ/cm$^2$. A, B and C label the regions of interest in squares. The color bars show the normalized diffraction intensity as used in Fig. 2c.

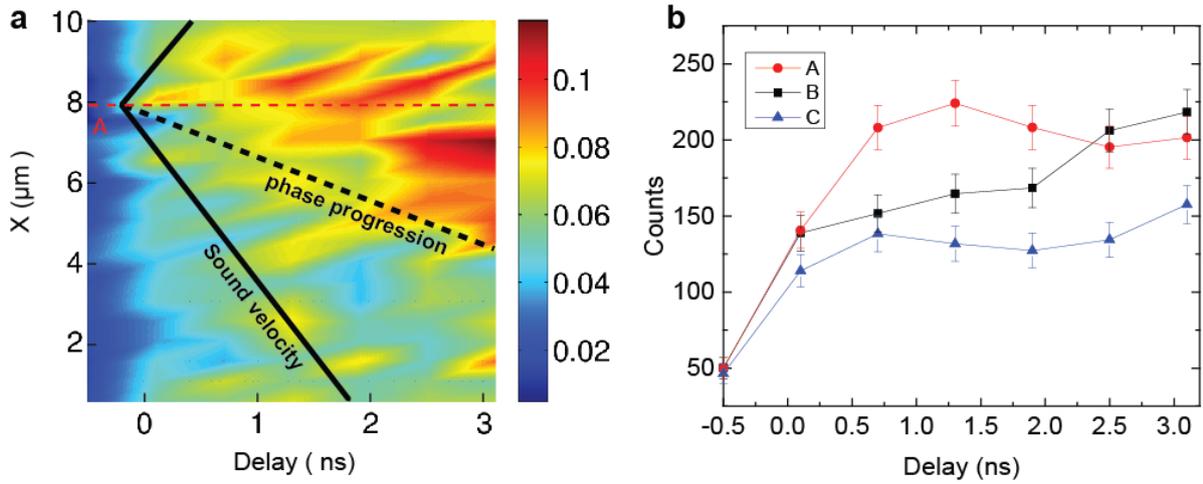

**Figure 4** (a) Space-time map of the diffracted intensities from the R phase, which is formed by the intensity line-cut along the red dotted line in Fig. 3 at various time delays. The black dashed line shows the averaged phase boundary defined roughly along 50% of the maximal intensity of the map. The color bar shows the normalized intensity as in Fig. 2c. Its slope is the speed of the in-plane phase propagation near site A. The solid black lines indicate the speed of sound in $VO_2$ and the red dashed line is a reference for non-propagating features. (b) The integrated intensity of three individual sites labeled as A, B and C in Fig. 3 as a function of time. Error bars correspond to the uncertainty associated with counting statistics.

# Supplemental Materials for "Mesoscopic structural phase progression in photo-excited VO$_2$ revealed by time-resolved x-ray diffraction microscopy"

*Yi Zhu[1], Zhonghou Cai[1], Pice Chen[2, †], Qingteng Zhang[2, §], Matthew J. Highland[3], Il Woong Jung[4], Donald A. Walko[1], Eric M. Dufresne[1], Jaewoo Jeong[5], Mahesh G. Samant[5], Stuart S. P. Parkin[5,6], John W. Freeland[1], Paul G. Evans[2], Haidan Wen[1*]*

[1] Advanced Photon Source, Argonne National Laboratory, Argonne, Illinois 60439, USA
[2] Department of Materials Science and Engineering, University of Wisconsin–Madison, Madison, Wisconsin 53706, USA
[3] Materials Science Division, Argonne National Laboratory, Argonne, Illinois 60439, USA
[4] Center for Nanoscale Materials, Argonne National Laboratory, Argonne, Illinois 60439, USA
[5] IBM Almaden Research Center, San Jose, California 95120, USA
[6] Max Planck Institute for Microstructure Physics, Halle 06120, Germany

*Corresponding author: wen@aps.anl.gov
† Present address: Department of Materials Science and Engineering, Northwestern University, Evanston, Illinois 60208, USA
§ Present address: Advanced Photon Source, Argonne National Laboratory, Lemont, Illinois 60439, USA

## 1. Technical details of the experimental setup

The laser pump, hard x-ray microdiffraction imaging probe experiment was carried out at the 7-ID-C beamline of the Advanced Photon Source (APS) (Fig. S1). An ultrafast Ti:Sapphire laser system (Micra and Legend from Coherent Inc.) provides laser pulses with 60 fs pulse duration and 1.55 eV photon energy at 1 kHz rate to photo-excite a thin film VO$_2$ sample grown on Al$_2$O$_3$ (001) substrate. The laser pulse is focused to ~ 200 μm (FWHM) in diameter on the sample surface.

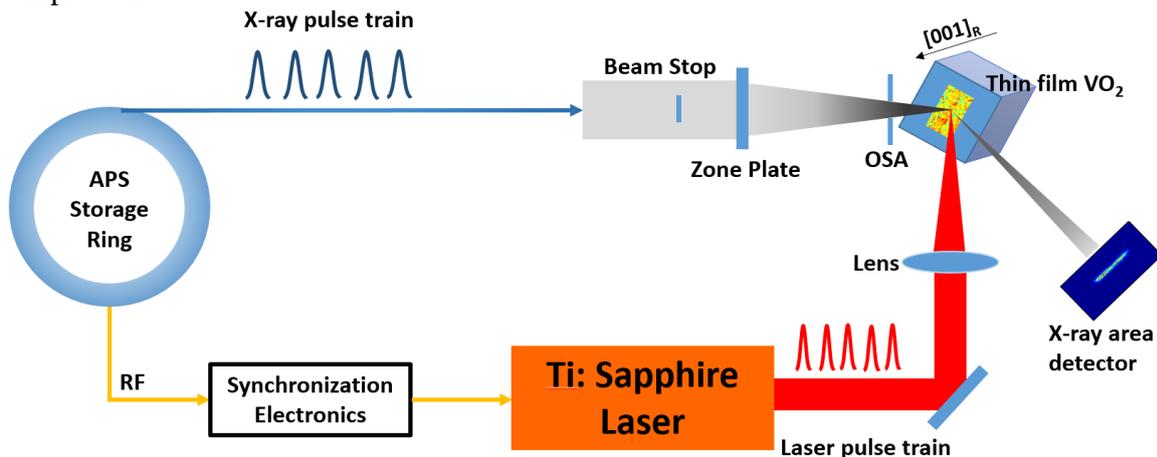

Figure S1 Experimental setup. A femtosecond laser pulse train is synchronized to an x-ray pulse train from the Advanced Photon Source (APS) storage ring. The x-ray pulses are focused at the sample position by a zone plate with a diameter of 160 μm and an out-zone width of 100 nm. The femtosecond laser pulses are focused onto the sample surface and overlapped with the x-ray probe beam. The x-rays horizontally diffracted from VO$_2$ are detected by an x-ray area detector.

Monochromatic 10 keV hard x-ray pulses with 0.1% band width from the APS synchrotron storage ring are focused by an x-ray Fresnel zone plate. The beam stop and the optical sorting aperture (OSA) are employed to block the central non-diffracted x-ray component and the higher order diffracted x-ray component respectively. The incident x-ray flux through the zone plate with a diameter of 160 μm is estimated to be ~$4\times10^{10}$ photons per second. At the sample position, the flux is reduced to $3.2\times10^9$ photons per second, in agreement with the 8% efficiency of the zone plate. The x-ray footprint on the sample is 350 nm and 830 nm (FWHM) in the vertical and horizontal directions respectively. Spatial overlap of the pump and probe beams is achieved by steering the laser beam to be concentered with the x-ray beam on the sample, aided by a micrometer-sized phosphor particle that fluoresces under the illumination of both optical and x-ray beam.

The ultrafast laser system is electronically synchronized to the RF (radio frequency) signal of the APS storage ring with a timing jitter less than 10 ps. The arrival time of laser pulses to the sample can be controlled electronically with respect to that of the x-ray pulses using a delay generator. The time resolution of the experiment is determined by the probe x-ray pulse duration of 100 ps.

The thin film $VO_2$ sample is mounted on a XYZ nano-positioning piezo stage (nPoint, Inc.) with the *c*-axis along the out-of-plane direction. The piezo stage is mounted to the inner circle of a two-circle diffractometer. During the experiment, a $10\times10$ μm$^2$ region is raster scanned using the piezo stage with a step size of 250 nm. This field of view is chosen so that the area is small enough to be homogenously excited by an optical pulse and big enough to illustrate the phase transition spatial features.

The x-ray diffraction peaks from the monoclinic (M) and rutile (R) phases are measured with an x-ray area detector (Pilatus 100K, DECTRIS Ltd.) mounted on the outer circle of the diffractometer. The repetition rate of the x-ray pulses is 6.5 MHz. Only the x-ray pulses that are paired with the excitation laser pulses are recorded by gating the detector at 1 kHz laser repetition rate, providing an effective time-resolved x-ray flux of ~$10^6$ photons per second at the sample position.

## 2. One-dimensional thermal transport

The out-of-plane phase progression as discussed in the main text shown in Fig. 2(c) is driven by thermal transport. This is due to inhomogeneous optical excitation along the out-of-plane direction as the film thickness is twice as thick as the penetration depth of the 800 nm excitation laser pulse. As a result, only surface layers of the $VO_2$ film can be sufficiently excited to the R phase at low excitation fluence, while the x-ray pulse is able to probe the film thoroughly in depth. The progression of this phase boundary deeper into the film results in the increase of R phase composition in the probed region following the excitation. To quantitatively study the dynamics associated with the out-of-plane phase progression, we use the intensity change of the Bragg diffraction peak of the M phase to gauge the fractional composition of the excited $VO_2$ along the depth of the probed region. The diffraction intensity at the fixed angle of 25.62 degree is corrected as the integrated intensity change of the Bragg peak based on the temperature-dependent XRD measurements. Using this intensity-to-thickness calibration, at 100 ps ($t_1$) after optical excitation with a pump laser fluence of 14 mJ/cm$^2$ (shown by the purple curve), we find that the M phase decreased by 14%, corresponding to 28 nm-thick $VO_2$ that is transformed into the R phase. The thermalization of the film leads to a reduction of thermal gradient at the phase boundary. The progression eventually loses the driving force (thermal gradient) for the out-of-plane direction at $t_2$, which is defined as the time when the fast intensity reduction transits into a slower linear reduction, shown as the dashed horizontal lines in Fig. S2. As the diffraction intensity of the M phase further decreases by 36% at $t_2$, the thickness of R phase increases from 28 nm to 73 nm (Fig. S2). At a higher pump fluence of 24 mJ/cm$^2$, thicker layer of the film is converted from M to R phase after the optical excitation, which results in a larger initial diffraction intensity change at 100 ps. The out-of-plane progression ends at later time $t_3$ with a phase boundary progression from 86 to 132 nm. The time scales of the out-of-plane phase progression agree with a thermal diffusion process. At both pump fluences, the phase progression over a length of $x \approx 50$ nm is about 1 ns. This is consistent with the thermal diffusion time of $t \approx \frac{x^2}{D} = 1.25$ ns, where $D=0.02$ cm$^2$/s is the thermal diffusion coefficient of VO$_2$ [S1].

The magnitude of the initial change of the diffraction intensity at 100 ps is proportional to the density of the nucleation sites that can transform upon optical excitation. At lower pump fluence, fewer sites in the illuminated volume of VO$_2$ will be converted directly by absorption of photons and yield smaller number of pinning sites, as seen by the smaller change at $t_1$ in the pink and blue curves comparing to the red and black curves in Fig. S2. As the pump fluence increases, the number of sites that can be excited across the phase transition also increases, leading to larger magnitude of the intensity change.

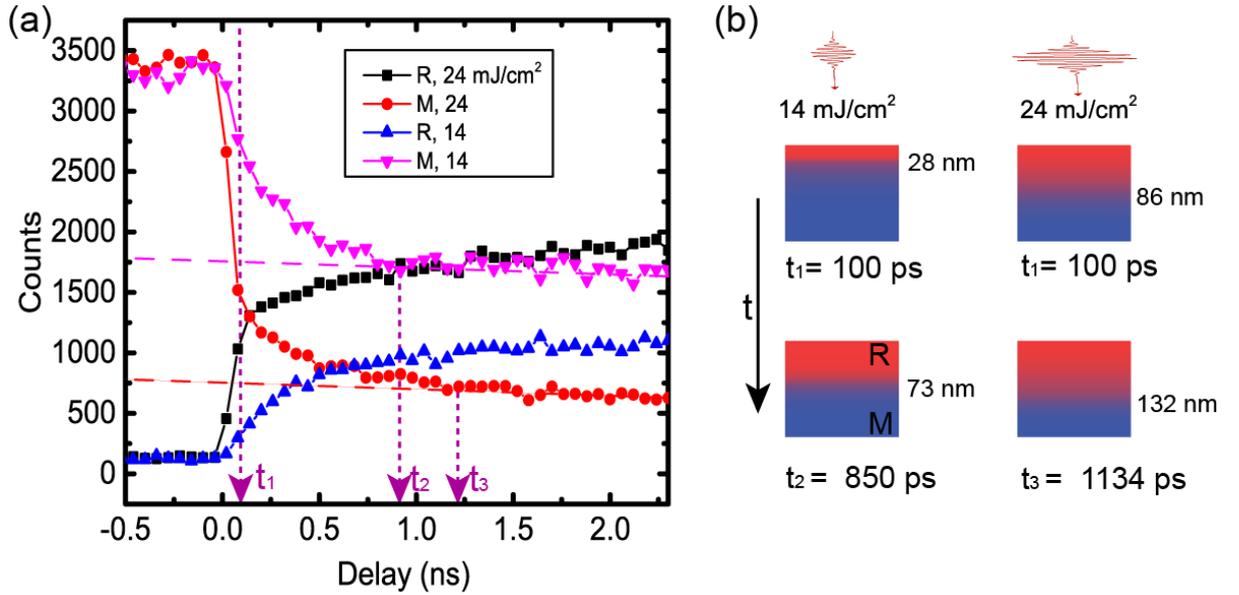

Figure S2 (a) Diffraction intensity of M and R phases as a function of time excited at two optical incident fluences. The purple dashed arrows show the start ($t_1$=100 ps, limited by x-ray pulse duration) and end ($t_2$, $t_3$) time of the longitudinal phase progression. The dashed horizontal lines are linear fits for the slow components on several ns time scales. (b) The schematics of the out-of-plane phase progression at two pump fluences. The red and blue regions indicate the layers of R and M lattice.

The change of diffraction intensity within ~1 ns can be fitted by a one-dimensional thermal transport model (Fig. S3). The thermal transport simulation was based on the VO$_2$/Al$_2$O$_3$ heterostructure using similar approach in Ref. [S2]. The initial temperature profile of VO$_2$ film is exponentially dependent on the position $z$ along the depth direction:
$$T(0 < z < D, t = 0) = T_0 \exp(-\alpha z),$$
where α is the absorption coefficient of 800 nm light. The temperature profile as a function of time $t$ was obtained by numerically solving the one-dimensional heat equation. As heat flows from the surface toward deeper VO$_2$ layers, larger volume of VO$_2$ is heated above the transition temperature, driving the progression of the R-M phase boundary into deep layers. The R-M phase boundary is defined by the layer of VO$_2$ at the transition temperature of 338 K. We track the position of R-M phase boundary and plot its change as a function of time in Fig. S3. It is in agreement with the measurements of the R phase diffraction intensity change.

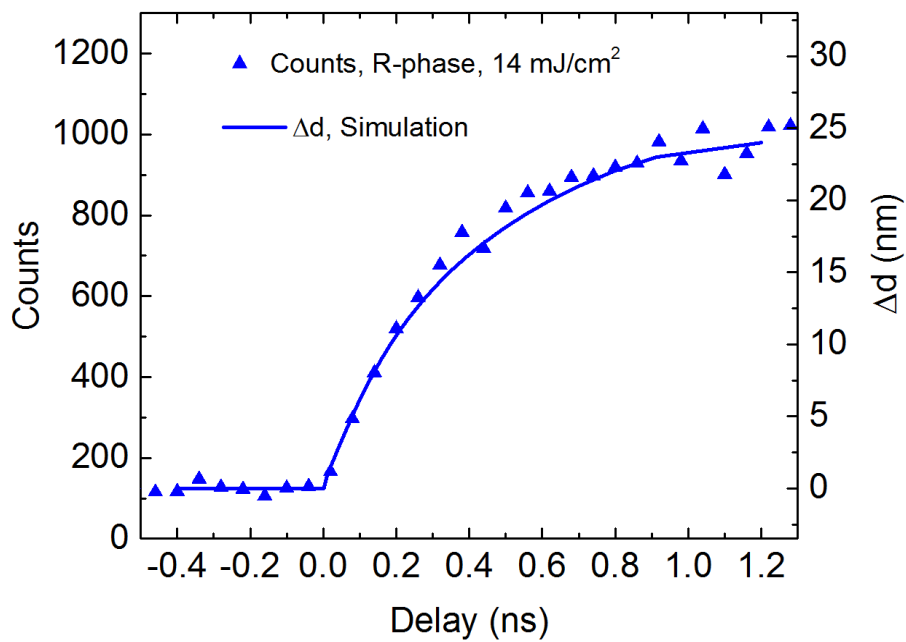

Figure S3 The diffraction intensity of R phase as a function of time is compared with the simulation result. The solid curve shows the change of R phase thickness as a function of time, driven by thermal transport along the out-of-plane direction.